\documentclass[11pt]{article}

\input epsf

\catcode`\@=11
\def\@citex[#1]#2{%
\if@filesw \immediate \write \@auxout {\string \citation {#2}}\fi
\@tempcntb\m@ne \let\@h@ld\relax \def\@citea{}%
\@cite{%
  \@for \@citeb:=#2\do {%
    \@ifundefined {b@\@citeb}%
      {\@h@ld\@citea\@tempcntb\m@ne{\bf ?}%
      \@warning {Citation `\@citeb ' on page \thepage \space undefined}}%
      {\@tempcnta\@tempcntb \advance\@tempcnta\@ne%
      \@tempcntb\number\csname b@\@citeb \endcsname \relax%
      \ifnum\@tempcnta=\@tempcntb 
        \ifx\@h@ld\relax%
          \edef \@h@ld{\@citea\csname b@\@citeb\endcsname}%
        \else%
          \edef\@h@ld{\ifmmode{-}\else--\fi\csname b@\@citeb\endcsname}%
        \fi%
      \else
        \@h@ld\@citea\csname b@\@citeb \endcsname%
        \let\@h@ld\relax%
      \fi}%
    \def\@citea{,\penalty\@highpenalty\,}%
  }\@h@ld
}{#1}}

\def\@citeb#1#2{{[#1]\if@tempswa , #2\fi}}
\def\@citeu#1#2{{$^{#1}$\if@tempswa , #2\fi }}
\def\@citep#1#2{{#1\if@tempswa , #2\fi}}

\def\bcites{         
        \catcode`\@=11
        \let\@cite=\@citeb
        \catcode`\@=12
}

\def\upcites{         
        \catcode`\@=11
        \let\@cite=\@citeu
        \catcode`\@=12
}

\def\plaincites{      
        \catcode`\@=11
        \let\@cite=\@citep
        \catcode`\@=12
}

\newcount\hour
\newcount\minute
\newtoks\amorpm
\hour=\time\divide\hour by 60
\minute=\time{\multiply\hour by 60 \global\advance\minute by-\hour}
\edef\standardtime{{\ifnum\hour<12 \global\amorpm={am}%
        \else\global\amorpm={pm}\advance\hour by-12 \fi
        \ifnum\hour=0 \hour=12 \fi
        \number\hour:\ifnum\minute<10 0\fi\number\minute\the\amorpm}}
\edef\militarytime{\number\hour:\ifnum\minute<10 0\fi\number\minute}

\def\draftlabel#1{{\@bsphack\if@filesw {\let\thepage\relax
   \xdef\@gtempa{\write\@auxout{\string
      \newlabel{#1}{{\@currentlabel}{\thepage}}}}}\@gtempa
   \if@nobreak \ifvmode\nobreak\fi\fi\fi\@esphack}
        \gdef\@eqnlabel{#1}}
\def\@eqnlabel{}
\def\@vacuum{}
\def\marginnote#1{}
\def\draftmarginnote#1{\marginpar{\raggedright\scriptsize\tt#1}}
\def\draft{
        \pagestyle{plain}
        \overfullrule=2pt
        \oddsidemargin -.5truein
        \def\@oddhead{\sl \phantom{\today\quad\militarytime} \hfil
        \smash{\Large\sl DRAFT} \hfil \today\quad\militarytime}
        \let\@evenhead\@oddhead
        \let\label=\draftlabel
        \let\marginnote=\draftmarginnote
        \def\ps@empty{\let\@mkboth\@gobbletwo
        \def\@oddfoot{\hfil \smash{\Large\sl DRAFT} \hfil}
        \let\@evenfoot\@oddhead}
        \def\@eqnnum{(\theequation)\rlap{\kern\marginparsep\tt\@eqnlabel}%
        \global\let\@eqnlabel\@vacuum}  }

\def\blackfonts{
        \font\blackboard=msbm10 scaled\magstep1
        \font\blackboards=msbm8
        \font\blackboardss=msbm6
}

\def\prep{         
        \catcode`\@=11
        \input art10.sty
        \catcode`\@=12
        
        \let\small\null
        \def\blackfonts{
                \font\blackboard=msbm10
                \font\blackboards=msbm7
                \font\blackboardss=msbm5
        }
        \let\sl\it
        \twocolumn
        \sloppy
        \voffset=-2.54truecm
        \hoffset=-2.54truecm
        \flushbottom
        \parindent 1em
        \leftmargini 2em
        \leftmarginv .5em
        \leftmarginvi .5em
        \marginparwidth 48pt
        \marginparsep 10pt
        \setlength{\columnsep}{2truecm}
        \setlength{\textwidth}{25.4truecm}
        \setlength{\textheight}{17truecm}
        \baselineskip=16pt
        \oddsidemargin .18truein
        \evensidemargin .17truein
}

\def\eqalign#1{\null\,\vcenter{\openup\jot\m@th
  \ialign{\strut\hfil$\displaystyle{##}$&$\displaystyle{{}##}$\hfil
      \crcr#1\crcr}}\,}
\def\eqalignno#1{\displ@y \tabskip\centering
  \halign to\displaywidth{\hfil$\@lign\displaystyle{##}$\tabskip\z@skip
    &$\@lign\displaystyle{{}##}$\hfil\tabskip\centering
    &\llap{$\@lign##$}\tabskip\z@skip\crcr
    #1\crcr}}

\def\section{\@startsection {section}{1}{\z@}{3.ex plus 1ex minus
 .2ex}{2.ex plus .2ex}{\large\bf}}
\def\subsection{\@startsection{subsection}{2}{\z@}{2.75ex plus 1ex minus
 .2ex}{1.5ex plus .2ex}{\bf}}        

\def\appendix{{\newpage\section*{Appendix}}\let\appendix\section%
        {\setcounter{section}{0}
        \gdef\thesection{\Alph{section}}}\section}

\def\abstract{\if@twocolumn
\section*{Abstract}
\else 
\begin{center}
{\bf Abstract\vspace{-.5em}\vspace{0pt}}
\end{center}
\quotation
\fi}

\catcode`\@=12

\def\sqr#1#2{{\vcenter{\vbox{\hrule height.#2pt\hbox{\vrule width.#2pt 
height#1pt \kern#1pt \vrule width.#2pt}\hrule height.#2pt}}}}

\def\=d{\,{\buildrel\rm def\over =}\,}

\def\i3p{\p32\int d^3p}

\def\As{A\hbox to 1pt{\hss /}}
\def\np4{\int d^4p_1\cdots d^4p_{n-1}\, }

\def\nx4{\int d^4x_1\ldots d^4x_n\, }

\def\kon#1#2{\vbox{\halign{##&&##\cr
\lower4pt\hbox{$\scriptscriptstyle\vert$}\hrulefill &
\hrulefill\lower4pt\hbox{$\scriptscriptstyle\vert$}\cr $#1$&
$#2$\cr}}}

\def\konv#1#2#3{\hbox{\vrule height12pt depth-1pt}
\vbox{\hrule height12pt width#1cm depth-11.6pt}
\hbox{\vrule height6.5pt depth-0.5pt}
\vbox{\hrule height11pt width#2cm depth-10.6pt\kern5pt
      \hrule height6.5pt width#2cm depth-6.1pt}
\hbox{\vrule height12pt depth-1pt}
\vbox{\hrule height6.5pt width#3cm depth-6.1pt}
\hbox{\vrule height6.5pt depth-0.5pt}}
\def\konu#1#2#3{\hbox{\vrule height12pt depth-1pt}
\vbox{\hrule height1pt width#1cm depth-0.6pt}
\hbox{\vrule height12pt depth-6.5pt}
\vbox{\hrule height6pt width#2cm depth-5.6pt\kern5pt
      \hrule height1pt width#2cm depth-0.6pt}
\hbox{\vrule height12pt depth-6.5pt}
\vbox{\hrule height1pt width#3cm depth-0.6pt}
\hbox{\vrule height12pt depth-1pt}}

\def\konw#1#2#3{\hbox{\vrule height12pt depth-1pt}
\vbox{\hrule height12pt width#1cm depth-11.6pt}
\hbox{\vrule height6.5pt depth-0.5pt}
\vbox{\hrule height12pt width#2cm depth-11.6pt \kern5pt
      \hrule height6.5pt width#2cm depth-6.1pt}
\hbox{\vrule height6.5pt depth-0.5pt}
\vbox{\hrule height12pt width#3cm depth-11.6pt}
\hbox{\vrule height12pt depth-1pt}}

\def\i{{\rm int}}

\def\m3{{\mu_1\mu_2\mu_3}}

\def\p{{(+)}}

\def\be{\begin{equation}}       \def\eq{\begin{equation}}
\def\ee{\end{equation}}         \def\eqe{\end{equation}}

\def\bea{\begin{eqnarray}}      \def\eqa{\begin{eqnarray}}
\def\ena{\end{eqnarray}}        \def\eea{\end{eqnarray}}
                                \def\eqae{\end{eqnarray}}

\def\ba{\begin{array}}
\def\ea{\end{array}}
\def\unit{1 \hskip-.3em \raise2pt\hbox{$ \scriptstyle |$ } }

\def\i{\iota}

\def\m{\mu}

\def\p{\pi}                
\def\t{\tau}

\def\bop#1{\setbox0=\hbox{$#1M$}\mkern1.5mu
        \vbox{\hrule height0pt depth.04\ht0
        \hbox{\vrule width.04\ht0 height.9\ht0 \kern.9\ht0
        \vrule width.04\ht0}\hrule height.04\ht0}\mkern1.5mu}

\def\>{\rangle} 

\def\<{\langle} 
\def\Dsl{D \hskip-.6em \raise1pt\hbox{$ / $ } }

\def\sl#1{\rlap{\hbox{$\mskip 1 mu /$}}#1}
\def\leftrightarrowfill{$\mathsurround=0pt \mathord\leftarrow \mkern-6mu
       \cleaders\hbox{$\mkern-2mu \mathord- \mkern-2mu$}\hfill
       \mkern-6mu \mathord\rightarrow$}
\def\dvec#1{\vbox{\ialign{##\crcr
       \leftrightarrowfill\crcr\noalign{\kern-1pt\nointerlineskip}
       $\hfil\displaystyle{#1}\hfil$\crcr}}}          
\def\hook#1{{\vrule height#1pt width0.4pt depth0pt}}
\def\leftrighthookfill#1{$\mathsurround=0pt \mathord\hook#1
       \hrulefill\mathord\hook#1$}
\def\underhook#1{\vtop{\ialign{##\crcr                 
       $\hfil\displaystyle{#1}\hfil$\crcr
       \noalign{\kern-1pt\nointerlineskip\vskip2pt}
       \leftrighthookfill5\crcr}}}
\def\smallunderhook#1{\vtop{\ialign{##\crcr      
       $\hfil\scriptstyle{#1}\hfil$\crcr
       \noalign{\kern-1pt\nointerlineskip\vskip2pt}
       \leftrighthookfill3\crcr}}}

\def\sfrac#1#2{{\vphantom1\smash{\lower.5ex\hbox{\small$#1$}}\over
       \vphantom1\smash{\raise.4ex\hbox{\small$#2$}}}} 
\def\bfrac#1#2{{\vphantom1\smash{\lower.5ex\hbox{$#1$}}\over
       \vphantom1\smash{\raise.3ex\hbox{$#2$}}}}      
\def\afrac#1#2{{\vphantom1\smash{\lower.5ex\hbox{$#1$}}\over#2}}  
\def\on#1#2{{\buildrel{\mkern2.5mu#1\mkern-2.5mu}\over{#2}}}
\def\ddt#1{\on{\hbox{\LARGE .\kern-2pt.}}#1}             
\def\tdt#1{\on{\hbox{\LARGE .\kern-2pt.\kern-2pt.}}#1}   

\def\boxes#1{
       \newcount\num
       \num=1
       \newdimen\downsy
       \downsy=-1.5ex
       \mskip-2.8mu
       \bo
       \loop
       \ifnum\num<#1
       \llap{\raise\num\downsy\hbox{$\bo$}}
       \advance\num by1
       \repeat}
\def\boxup#1#2{\newcount\numup
       \numup=#1
       \advance\numup by-1
       \newdimen\upsy
       \upsy=.75ex
       \mskip2.8mu
       \raise\numup\upsy\hbox{$#2$}}

\newskip\humongous \humongous=0pt plus 1000pt minus 1000pt
\def\caja{\mathsurround=0pt}
\def\eqalign#1{\,\vcenter{\openup2\jot \caja
       \ialign{\strut \hfil$\displaystyle{##}$&$
       \displaystyle{{}##}$\hfil\crcr#1\crcr}}\,}
\newif\ifdtup

\def\1ov4{{1\over 4}}

\def\ddt{\dot{\t}}

\def\bfx{{\bf x}}

\newcommand{\rmd}{{\rm d}}
\newcommand{\rmi}{{\rm i}}

\newcommand{\beq}{\begin{equation}}
\newcommand{\eeq}{\end{equation}}
\def\ba{\begin{eqnarray}}
\def\ea{\end{eqnarray}}

\newcommand{\llr}{\langle\lambda\rangle}
\newcommand{\lthetar}{\langle\theta\rangle}

\begin{document}

\vskip-10pt \hfill CBPF-NF-055/99 \vskip-10pt \hfill KL-TH 99/02
\vskip-10pt \hfill THES-TP 99/13
\vskip-10pt \hfill {\tt
hep-th/9911177} \vskip0.3truecm
\begin{center}
\vskip 1.5truecm
{\Large\bf
Thermodynamics of the 2+1 dimensional Gross-Neveu model with
complex chemical potential}
\vskip 1.2truecm
{\bf H. R. Christiansen${}^{\dagger\diamond}$\footnote{e-mail:
{\tt hugo@cbpf.br} }, 
A. C. Petkou${}^{\star}$\footnote{e-mail:{\tt
    petkou@physik.uni-kl.de, Alexander von Humboldt Fellow}} ,
M. B. Silva Neto${}^{\dagger\ddag}$\footnote{
e-mail:{\tt sneto@if.ufrj.br} (At $^\ddag$ after October 1999)},
N. D. Vlachos$^{\star\star}$\footnote{e-mail:{\tt
    vlachos@physics.auth.gr}}}
\vskip .8truecm
${}^{\dagger}$ {\it Centro Brasileiro de Pesquisas F\'{i}sicas,
Rua Dr. Xavier Sigaud 150 - Urca,\\ CEP:22290-180, Rio de Janeiro,
Brazil.}

${}^\star$ {\it Fachbereich Physik, Theoretische Physik, Universit\"at
Kaiserslautern \\ Postfach 3049, 67653 Kaiserslautern, Germany.}

${}^\ddag$ {\it Instituto de F\'isica, Universidade Federal do Rio de
Janeiro, \\ Caixa Postal 68528, CEP:21945-970, Rio de Janeiro,
Brazil.}

${}^{\star\star}$ {\it Department of Theoretical Physics,
Aristotle University of Thessaloniki, \\
Thessaloniki 54006, Greece.}

${}^{\diamond}$ {\it Group of Theoretical Physics, Universidade 
Cat\'olica de Petr\'opolis,\\ Rua Bar\~ao de Amazonas 124,  
Petr\'opolis 25685-070, Rio de Janeiro, Brazil.}

\vskip 0.5truecm
\vskip 1truemm
\end{center}
\vskip 0.5truecm
\noindent{\bf Abstract:}

We study the thermodynamics of the 2+1 dimensional Gross-Neveu
model by introducing a
representation for the canonical partition function which
encodes both real and imaginary chemical potential cases. It is
pointed out that the latter case probes the thermodynamics of 
possible anyon-like excitations in the spectrum. It is also
intimately connected to the breaking of the discrete $Z$-symmetry
of the model, which we interpret as signaling anyon
deconfinement. Finally, the chiral properties of the model in the
presence of an imaginary chemical potential are discussed and
analytical results for the free-energy density at the transition
points are presented.

\vfill
\vskip4pt

\eject
\newpage

\newpage


\section{Introduction}

Chiral symmetry and its realization at finite temperature have been
matter of intense studies for a long time. One of the most
extensively used laboratories for this purpose has been the
Gross-Neveu model \cite{Gross-Neveu} both at zero and finite
temperature \cite{Rosenstein}\cite{Gross-Pisarski-Jaffe}.  
In the large-$N$ limit the
$2+1$ dimensional version of this theory is renormalizable and
provides a convenient arena for the survey of chiral symmetry;
 for instance, its dynamical breakdown at zero temperature and
its high temperature restoration. Recently,
there has been also considerable interest in studying the properties 
of chiral symmetry at finite density, or equivalently in the presence
of a chemical potential \cite{Hands,Kogut,Strouthos}. This is 
believed to be relevant to the understanding of the physics of hot 
and dense matter which is expected  to  be probed in the laboratory 
by the RHIC experiments \cite{Wilczek}.

In this work, we address the thermodynamics and chiral properties
of the 2+1 dimensional Gross-Neveu model in the {\it canonical}
formalism. The use of the  canonical formalism follows
naturally when one introduces a constraint on the fermion number.
Our discussion, however, will be based on a generalized
representation of the canonical partition function that
encodes the presence of a real chemical potential as a special
case. It is shown that the approach yields similar results to the 
grand canonical formalism outcomes, the latter
being the natural framework for studying systems in
the presence of a chemical potential \cite{Kapusta,Redlich}. This
generalized partition function will also be studied for imaginary
values of the chemical potential where it corresponds, rather
surprisingly, to a real free-energy density. \footnote{The study
of the thermodynamics of fermionic systems in the presence of an
imaginary chemical potential has been considered in the past
\cite{Weiss,Karch,Alford}. Such studies, however, have been
primarily focused on the technical virtues of the imaginary
chemical potential formalism in lattice simulations.}

We argue that the presence of an imaginary chemical potential is
intimately connected to the possibility of having anyon-like
excitations in the spectrum of the theory. We also show that the
imaginary chemical potential emerges naturally in 
 the 2+1 dimensional Gross-Neveu $U(1)$ gauged
model at finite temperature. Such a theory has infinitely many
$Z$-vacua \cite{Semenoff1} around which the gauge field may
fluctuate. These fluctuations have been shown to be connected to
peculiar excitations whose charge {\it is not} an integer multiple
of the elementary charge of the theory. We interpret these
excitations as being anyon-like \cite{Huang} and argue that our
generalized  canonical partition function gives their
free-energy density. As the gauge field fluctuations around the
$Z$-vacua lead to the spontaneous breakdown of the discrete
$Z$-symmetry, it appears that the latter is tied to an {\it anyon
confinement/deconfinement} transition. Finally, we study the
chiral properties of the Gross-Neveu model in the presence of an
imaginary chemical potential. In this case the theory is chirally
symmetric at high enough temperatures but it appears to be
unstable at $T=0$. For any non-zero value of the imaginary
chemical potential, chiral symmetry is broken at a certain
temperature  as the system cools down. This corresponds to a
second-order phase transition where the free-energy density of the
system is given by a remarkably simple analytic expression. There
also exists a particular value of the temperature at which the system 
appears to be in a bosonic phase.

The paper is organized as follows. In Sec.2 we briefly review the
2+1 dimensional Gross-Neveu model at finite temperature. In Sec.3
we set the stage for studying the model with a constraint on the
fermion number. Using a generalization of the well-known formula
which gives the canonical partition function \cite{Weiss,Karch,Alford} 
we derive analytic expressions for the
free-energy density, the fermion number density and the chiral
order parameter. Our formalism coincides with the standard 
grand-canonical approach when the chemical potential takes real
values. In Sec.4 we study the model in the presence of an
imaginary chemical potential and show that it corresponds to a
system  possessing anyon-like excitations. We establish a
connection  with the breakdown of the $Z$-symmetry
\cite{Semenoff1} of the $U(1)$ gauged Gross-Neveu model at finite 
temperature. In this way we give a
physical interpretation to the canonical partition function
with imaginary chemical potential as representing the free-energy
needed to immerse an excitation of imaginary charge in the
spectrum. We argue that such excitations are anyon-like. Next, we
present results related to the properties of chiral symmetry in
the presence of an imaginary chemical potential at finite
temperature. In Sec.5 we summarize and discuss possible
implications of our results.


\section{The 2+1 dimensional Gross-Neveu model at finite temperature}

The Euclidean Gross-Neveu model may be  defined
by the Lagrangian
density \footnote{For the
Euclidean gamma matrices we use the Hermitean representation
$\gamma_i=\left( \begin{array}{cc} \sigma^i & 0 \\ 0 & -\sigma^i
\end{array}\right)$
with $\sigma^i$, $i=1,2,3$ the usual Pauli matrices and
$\gamma_0\equiv\gamma_3$. In this reducible
representation there exist {\it two} $\gamma_5$-like  matrices $\left(
\begin{array}{cc} 0 & 1 \\ 1 & 0
\end{array}\right)$ and $\left( \begin{array}{cc} 0 & 1 \\ -1 & 0
\end{array}\right)$.}
\cite{Rosenstein,Zinn-Justin}
\beq
{\cal L}=-\bar{\psi}^{i}\partial\!\!\!\slash\psi^i-
\frac{g}{2}(\bar{\psi}^i\psi^i)^{2},
\label{GN-Model}
\eeq
where $\bar{\psi}^i$, $\psi^i$, $i=1,2,..,N$,  are four-component
Dirac fermions and $g$ is the coupling. In $d=3$  the massless
model above is $U(2N)$-invariant and possesses a $Z_2$ discrete
``chiral'' symmetry
\bea
\psi\rightarrow\gamma_{5}\psi\,,\,\,\,\,\, \bar{\psi}\rightarrow
-\gamma_5\bar{\psi}.
\eea
This  model has been extensively used as a testing ground for
studying the mechanism of chiral symmetry breaking in  QCD
\cite{Strouthos}.

For large-$N$, the model is studied in a $1/N$ expansion where it
is renormalizable \cite{Zinn-Justin}. One introduces an auxiliary
scalar field $\lambda(x)$ and integrates the fermions out. The
partition function (generating functional) reads
\bea {\cal Z}&=&\int({\cal D}\lambda) e^{-NI_{eff}(\lambda,G)}\,,
\label{Original-Partition-Function}\\ I_{eff}(\lambda,G)&=&
\frac{1}{2 G}\int\rmd^{3}x{\;}\lambda^{2}(x)-
Tr\left[\ln(\partial\!\!\!\slash+\lambda)\right]\,,
\label{Original-Effective-Action} \eea
where the rescaled coupling $G=gN$ is kept finite as
$N\rightarrow \infty$.

To leading order in $1/N$  the renormalized theory manifests
itself in two different phases distinguished by a zero or a
non-zero expectation value of $\langle\lambda\rangle$. This
manifestation depends on the value of the renormalized coupling
$1/G_r$ as compared to the critical coupling $1/G_*
=4(2\pi)^{-3}\int \rmd^3 p/p^2$. For $1/G_r > 1/G_*$,
$\langle\lambda\rangle=0$ and the theory is in a weakly-coupled
phase where chiral symmetry is unbroken. For $1/G_r <1/G_*$,
$\langle\lambda\rangle\neq 0$ and the theory is in a
strongly-coupled phase where chiral symmetry is broken. In the
latter case the fermions acquire a mass proportional to $\llr$.
Clearly, $\llr$ plays the role of an order parameter for the
chiral phase transition.

From the Euclidean formulation above one straightforwardly
switches over to thermodynamics at temperature $T=1/\beta$  by
making $x_0$ finite with  length $L=1/T$ \footnote{In the
following we shall invariably use $T$, $\beta$ and $L$ bearing in
mind the relations $\beta=L=1/T$.} and imposing periodic
(antiperiodic) boundary conditions over the interval $[0,L]$ for
bosonic (fermionic) variables. In this way the bulk theory in 2+1
dimensions corresponds to a two-dimensional quantum system at zero
temperature. It then follows that the $T=0$ system can be
``prepared'', by appropriately tuning the coupling constant, to be
either in the chirally symmetric or in the chirally broken phase.
Had it been ``prepared'' to be in the broken (ordered) phase, one
would expect that there exists a high temperature phase transition
to the symmetric (disordered) phase. Such a transition is allowed
(not forbidden) by the Mermin-Wagner-Coleman theorem as the relevant 
symmetry is discrete. \footnote{Compare this situation to the 2+1
dimensional $O(N)$ vector model \cite{Rosenstein2,Tassos1}. In
that case, the bulk theory again corresponds to a two-dimensional
quantum system at $T=0$ which can be in {\it either} an
$O(N)$-symmetric phase {\it or} an $O(N)$-broken phase.
However, the $T>0$ theory can only be in the $O(N)$-symmetric
phase as the relevant $O(N)$ symmetry is continuous and cannot be
broken in $d=2$.}

To study the chiral symmetry restoration one calculates the
partition function
(\ref{Original-Partition-Function}) by the steepest descent
method for large-$N$. This amounts to performing a $1/N$ expansion for
${\cal Z}$
around its saddle-points, which  correspond to uniform values of
$\llr$. The latter are obtained from the gap equation
\beq \frac{1}{G}=\frac{4}{L}\sum_{n=-\infty}^{\infty}
\int\frac{\rmd^{2}{\bf p}}{(2\pi)^{2}} \frac{1}{{\bf
p}^{2}+\omega_{n}^{2}+\llr^{2}}\,,\,\,\,\,\, \omega_n
=(2n+1)\frac{\pi}{L}\,,n=0,\pm 1,\pm2,\dots \,.
\label{Gap-Equation} \eeq
This expression is divergent but all UV divergences at finite
temperature are the same as at zero temperature
\cite{Zinn-Justin}. Therefore, one can renormalize
(\ref{Gap-Equation}) by substituting  for $1/G$ its corresponding
renormalized value at $T=0$. For the system to be in the chirally
broken phase at $T=0$ one sets
\beq
\frac{1}{G}\rightarrow \frac{1}{G_r}=
-\frac{m}{\pi},
\label{Bulk-Renormalized-Coupling}
\eeq
where $m>0$ is the dynamically induced mass of the
elementary fermions at $T=0$. Then, from (\ref{Gap-Equation}) one obtains
\beq
\llr=\frac{2}{L}
\ln{\left[
\frac{e^{L m/2}+\sqrt{e^{L m}-4}}{2}
\right]}\,,
\label{The-Gap}
\eeq
which gives  the dependence of the order parameter $\llr$ on the
temperature. In particular, $\llr$ vanishes at the second-order
phase transition point $1/L_c=T_{c}=m/2\ln{2}$ where chiral
symmetry is restored. At this point and for higher temperatures,
the free-energy density is given to leading-$N$ by \footnote{In
our calculations we always normalize the free-energy density in
such a way that it vanishes in the bulk, i.e. at T=0
\cite{Cardy}.}
\beq
\frac{f}{N}=-\frac{3\zeta(3)}{2\pi L^3}\,,
\label{f-N-Dirac}
\eeq
and coincides with the free-energy density of $2N$ massless
four-component Dirac fermions \cite{Tassos-Marcello}.


\section{Thermodynamics of the 2+1 dimensional Gross-Neveu model in
the canonical formalism}

\subsection{General Setting}

The canonical formalism for the analysis of the thermodynamics of a
system has been recently employed in studies of
fermionic systems at finite baryon density \cite{Alford,Karch}. The reason
is that it bypasses the
usual sign problem of the Euclidean fermion determinant for real
values of the chemical potential. Since the three-dimensional
Gross-Neveu model does not undergo this sign problem, it has
been extensively studied in the standard  grand-canonical
formalism  both analytically and numerically \cite{Hands,Kogut}.
Nevertheless, it would still be interesting to perform a direct
analysis  in the  canonical formalism in order to test
results obtained previously. Furthermore, in doing so, we obtain some
surprising new results which shed new light into the
thermodynamics of the model.

The  canonical partition function can be obtained as the
thermal average over eigenstates of the number operator
$\hat{N}=\int\rmd^2{\it {\bf x}}{\;}\psi^{\dagger}(\tau,{\bf
x})\psi(\tau,{\bf x})$ with fixed eigenvalues $B$. Namely,
\cite{Kapusta}
\beq
{\cal Z}(T,V,B) =Tr\left(e^{-\beta\hat{H}}\delta(\hat{N}-B)\right),
\label{Constrained-Part-Func}
\eeq
where $\hat{H}$ is the
Hermitian Hamiltonian. As usual, ${\hat{N}}$ measures the excess
of fermions over anti-fermions in the spectrum.

If one anticipates that for certain physical conditions the spectrum
contains free fermions or anti-fermions, then $B$ must be an
integer. This then leads to the following representation for the 
canonical partition function \cite{Weiss}

\beq
{\cal Z}(T,V,B)=\frac{1}{2\pi}\int_0^{2\pi}\rmd \phi
               \,e^{-\rmi \phi B} {\cal Z}(T,V,i\phi)\,,\label{IntegerB}
\eeq
where, ${\cal Z}(T,V,{\rm i}\phi)$ is the  grand-canonical
counterpart with imaginary chemical potential. If on the
other hand one is interested in physical situations where $B$ is
the mean fermion (anti-fermion) density, then $B$ is real, not
necessarily integer, and (\ref{IntegerB}) is no more  a valid
representation.

In this work we propose a generalization of Eq.(\ref{IntegerB}) which
is suitable for both analytic and numerical work. Our
representation includes both cases of integer and non-integer real
values of $B$. Furthermore, it leads to some unexpected new
results when $B$ is imaginary. Our representation follows from
Eq.(\ref{Constrained-Part-Func}) if we write the delta-function
constraint  with the help of an auxiliary Lagrange multiplier
scalar field  $\theta(\tau)$ as
\bea
{\cal Z}(T,V,B)&=&\int({\cal D}\lambda)({\cal D}\theta)
e^{-N{\cal I}_{eff}(\lambda,g;\theta,B)}\nonumber \\
& = & e^{-\beta F(T,V,B)},
\label{Fint}
\eea
which enforces the {\it averaged} fermion number constraint $\langle
\hat{N}\rangle =B$. Here, $F(T,V,B)$ is the free-energy and
${\cal I}_{eff}$ is the
following effective action
\beq {\cal I}_{eff}(\lambda,g;\theta,B) = \rmi
\frac{B}{N}\int_{0}^{L}\rmd \tau{\;}\theta(\tau)
+\frac{1}{2G}\int_{0}^{L}\rmd \tau \int\rmd^2 {\bf x}{\;}
\lambda^{2}(\tau,{\bf x}) -Tr[\ln(\slash\!\!\!\partial
+\lambda(\tau,{\bf x})+ \rmi \gamma_{0}\theta(\tau))]_{L}\,.
\label{New-Model} \eeq
To evaluate the free-energy in (\ref{Fint}) for large-$N$, we
expand $I_{eff}$ around its stationary points assuming constant
(translation invariant) $\langle\lambda\rangle$ and $\lthetar$
configurations. These satisfy the following set of saddle-point
equations \footnote{As discussed in
\cite{Tassos-Marcello,Marcello} the regulating parameter $\rho$ in
(\ref{Second-Saddle-Point-Eq}) is necessary in order to take care
of the fact that the $Tr[...]$ and $\sum[...]$ operations do not
commute.}
\bea
\frac{1}{G}&=&\frac{4}{L}\sum_{n=-\infty}^{\infty}
\int\frac{\rmd^{2}{\bf p}}{(2\pi)^{2}}
\frac{1}{{\bf p}^{2}+(\omega_{n}+\lthetar)^{2}+\llr^{2}},
\label{First-Saddle-Point-Eq}\\
\rmi\frac{B}{N}&=&\lim_{\rho\rightarrow 0}\frac{4}{L}
\int\frac{\rmd^{2}{\bf p}}{(2\pi)^{2}}\sum_{n=-\infty}^{\infty}
\frac{e^{\rmi\omega_{n}\rho}(\omega_{n}+\lthetar)}
{{\bf p}^{2}+(\omega_{n}+\lthetar)^{2}+\llr^{2}},
\label{Second-Saddle-Point-Eq}
\eea
where $\omega_n =(2n+1)\pi/L$, $n=0,\pm 1,\pm 2,..$.
\cite{Marcello,Tassos-Marcello}. We can subtract the UV
divergences in Eq.(\ref{First-Saddle-Point-Eq})
\cite{Tassos-Marcello} by adjusting the
renormalized coupling as in (\ref{Bulk-Renormalized-Coupling}).
At the same time the system at $T=0$ can be arranged so as
to break chiral symmetry.  
After some algebra we obtain
\bea
0& = &\llr-m+\frac{1}{L}\left[ \ln\left(1+e^{-L\llr-\rmi L\lthetar}\right) +
\ln\left(1+e^{-L\llr+\rmi L\lthetar}\right)\right], \label{gap1}\\
\frac{\tilde{B}}{N} & = & \frac{\llr}{2\pi
L}\left[\ln\left(\frac{1+e^{-L\llr+\rmi L\lthetar}}{1+e^{-L\llr-\rmi
L\lthetar}}\right) -Li_2\left(-e^{-L\llr+\rmi L\lthetar}\right)
+Li_2\left(-e^{-L\llr-\rmi L\lthetar}\right)\right],\label{gap2}
\eea
where $\tilde{B}=B/V$ is the fermion density. From
(\ref{Fint}) and (\ref{New-Model}) we can now calculate the
renormalized free-energy density $f=F/V$ to
leading-$N$ and obtain
\bea
\frac{1}{N}f(\beta,m;\llr,\tilde{B}) & = & \frac{\llr^3}{3\pi}
-\frac{m\llr^2}{2\pi} +2\rmi \tilde{B}\lthetar +
\frac{1}{6\pi L^{3}}\left[Li_3(-e^{-L\llr+\rmi L\lthetar} )
+Li_{3}(-e^{-L\llr -\rmi L\lthetar}) \right.\nonumber \\
& & \left.-\ln(e^{-L\llr})\left
( Li_{2}(-e^{-L\llr +\rmi L\lthetar})+
Li_2(-e^{-L\llr -\rmi L\lthetar})\right)\right].
\label{Constrained-Free-Energy}
\eea
The functions $Li_n(z)$ are the
standard polylogarithms \cite{Lewin}.

In principle, the free-energy density (\ref{Constrained-Free-Energy})
together with the gap equations (\ref{gap1}) and
(\ref{gap2}) are sufficient for studying the
thermodynamic properties of the Gross-Neveu model to
leading-$N$. Before we proceed, however, we observe that requiring the
free-energy density in (\ref{Constrained-Free-Energy}) to be real
we are naturally led to distinguish two cases. 
Namely, as the terms involving polylogarithms in
(\ref{Constrained-Free-Energy}) are real for both real and
imaginary values of $\lthetar$, we can have {\it either} {\bf 1)}
$\lthetar$=imaginary and $\tilde{B}$=real, {\it or} {\bf 2)}
$\lthetar$=real and $\tilde{B}$=imaginary. Clearly, ${\bf 1)}$
would correspond to the usual case of the Gross-Neveu model with
real chemical potential. Nevertheless, we will show that case
${\bf 2)}$ probes some interesting properties of the Gross-Neveu
model too.


\subsection{Relation to the grand-canonical formalism}

For $\lthetar$=imaginary (\ref{Constrained-Free-Energy}) is the free-energy
density of the Gross-Neveu model with ordinary (real) chemical potential.
Indeed, setting $\rmi\lthetar = \mu< m$ we find that (\ref{gap1})
and (\ref{gap2}) coincide with  the corresponding expressions for
the gap equation and the fermion density, the latter was obtained for
the first time
in \cite{Kogut}. Namely,
\bea
\llr & = & \frac{1}{L}\ln\left[\frac{e^{Lm}-2\cosh(L\mu)+\sqrt
{ (e^{Lm}-2\cosh(L\mu))^{2}-4}}{2}\right], \\
\frac{\tilde{B}}{N} & = & \frac{\llr}{2\pi L}\left[\ln\left(\frac{1 +
e^{-L\llr +L\mu}}{1+e^{-L\llr -L\mu}}\right) -Li_2\left(-e^{-L\llr
+L\mu}\right) +Li_2\left(-e^{-L\llr
-L\mu}\right)\right]\,.
\eea
The thermodynamic properties here are well-known. As the chemical
potential $\mu$  increases from zero, the critical temperature
$T_{c}$ for the chiral symmetry restoration decreases. This
critical temperature becomes zero at some critical value of the 
chemical potential
$\mu_{c}\equiv m$ where the chiral symmetry restoration is of
first order. One can draw a physical picture of the chiral symmetry
restoration in terms of overlapping composites. When the
temperature is increased, chiral condensates begin to overlap as
their radius grow up. At some critical point the system is
mainly composed of overlapping condensates which, as a result, 
are no longer the good basis for describing the thermodynamic 
properties of the
system and the fermionic constituents must be taken into account.
It is then reasonable to expect that by increasing the baryon
density, which amounts to increasing the density of 
chiral condensates,  a lower critical temperature would be needed
for the system to reach the critical point above. In Fig.1 we plot
the critical lines in the $\llr$-$\mu$ plane for various values of
$T$. In Fig.2 we plot $\llr$ vs $T$ for various values of $\mu$.
In Fig. 3 we  plot the fermion density $\tilde{B}/N$ vs $T$ for
various values of $\mu$. For $\mu>0$ the fermion density is
discontinuous at a critical temperature $T_{c}(\mu)$.


\section{Imaginary chemical potential}

\subsection{The connection with an anyon-like system (anyon
  confinement/deconfinement)}

Consider the interaction of the Gross-Neveu model (\ref{GN-Model})
with an external $U(1)$
gauge potential $A_{\mu}(x)$. The Euclidean Lagrangian
density reads
\beq
{\cal L}=-\bar{\psi^i}(\slash\!\!\!\partial -\rmi e
\slash\!\!\!\!A)\psi^i-\frac{g}{2}(\bar{\psi^i}\psi^i)^2,
\label{GN-A-Model}
\eeq
where $e$ is the electric charge. Let us consider a constant
potential $A_0\equiv\theta$ along the ``time'' direction. We can
imagine embedding  the model above into a 4-dimensional space. Due
to the finite length of the $x_0$ dimension and the antiperiodic
boundary conditions of the fermions along it, the system may  be
viewed as existing in a 3-dimensional hyper-cylinder whose
axis is the 4$^{th}$ (unobservable) dimension. The constant
potential $A_0$ may now be regarded as the ``vector'' potential
generated by a thin solenoid of magnetic flux $\Phi=\theta L$
along the axis of the hyper-cylinder. Such a picture corresponds
to fermions encircling  a thin solenoidal magnetic flux and one
might expect to encounter Aharonov-Bohm type phenomena
\cite{Huang}.

The potential in (\ref{GN-A-Model}) may be gauged away by the transformation
\beq \psi\rightarrow e^{\rmi e\theta\tau}\psi \,\, . \eeq
Such a transformation, however,  ``twists'' the antiperiodic boundary
conditions for the fermions unless
\beq \theta=\frac{2\pi}{eL}k\,,\,\,\,\,\, k=0 \pm 1,\pm 2,\dots
 \label{Flux-Quantization} \,\, . \eeq
The configurations with ``twisted'' boundary conditions may be viewed
as anyon-like
excitations \cite{Huang}. For instance, the ``quasi-particle''
propagator of the above model (\ref{GN-A-Model})
\beq
S(x_{0},{\bf x};A_{0},\llr)  =  (\slash\!\!\!\partial +{\rm
i}e\gamma_{0}A_{0} +\llr)S_{0}(x_{0},{\bf x};A_{0},\llr)\,,
\eeq
has the following
representation as a sum over non-trivial topological paths
\beq
S_{0}(x_{0},{\bf x};A_{0},\llr)  =  \sum_{l=-\infty}^{\infty} e^{{\rm
i}\pi l} \,e^{{\rm i}(l L+x_{0})eA_{0}}\Pi(x_{0}+lL,{\bf x};\llr)\,,
\eeq
where $\Pi(x_{0},\bf{x};\llr)$ are free {\it boson} propagators of mass
$\llr$. Furthermore, since in our generalized representation of
the partition function (\ref{Fint}) we imposed a constraint on the
{\it average} fermion number, we can have local charge density
fluctuations $b(\tau,{\bf x})$ which give rise to the so-called
statistical gauge field $a_{\mu}(\tau,{\bf x})$ as
\beq
b(\tau,{\bf x})=\psi^{\dagger}(\tau,{\bf x})\psi(\tau,{\bf x}) -\tilde{B} =
\partial_1 a_2(\tau,{\bf x})-\partial_2 a_1(\tau,{\bf
  x})=\epsilon^{0ij}\partial_ia_j(\tau,{\bf x}) \label{Charge-Flux}\,.
\eeq
This (non-dynamical) field would be responsible for the anyon dynamics  
\cite{Arovas}. Finally,
the partition function of the free theory i.e (\ref{GN-A-Model})
at $g=0$, has been shown to reproduce the standard anyon virial
coefficients \cite{Farina}.

The existence of the anyonic excitations above may be tied to a
discrete $Z$-symmetry of the 2+1 dimensional Gross-Neveu model
interacting with  a standard $U(1)$ gauge field $A_{\mu}(x)$ at
temperature $T$. The partition function in this case is
\beq
{\cal Z}=\int ({\cal D}A_{\mu})({\cal D}\psi)({\cal D}\bar{\psi})
e^{-\int_{T} \left[\frac{1}{4}F_{\mu\nu}F_{\mu\nu}
-\bar{\psi}^{i}(\slash\!\!\!\partial -{\rm
i}e\slash\!\!\!\!A)\psi^{i}-\frac{g}{2}(\bar{\psi}^{i} \psi^{i})^{2}
\right]},
\label{U(1)-Plus-GN}
\eeq
where $F_{\mu\nu}=\partial_{[\mu}A_{\nu]}$ as usual. This theory
is invariant under 
\bea
A_{\mu}(\tau,{\bf x}) &\rightarrow&  A_{\mu}(\tau,{\bf
x})+\partial_{\mu}\chi(\tau,{\bf x})\,, \label{A-transf}\\
\psi(\tau,{\bf x})  &\rightarrow&  e^{\rmi  e\chi(\tau,{\bf
x})}\psi(\tau,{\bf x}) \,,\label{psi-transf}\\
\bar{\psi}(\tau,{\bf x})
&\rightarrow& \bar{\psi}(\tau,{\bf x})e^{-\rmi  e\chi(\tau,{\bf
x})}\,,\label{bpsi-transf}
\eea
which are the usual small gauge transformations provided they are
periodic in Euclidean time $\tau$
\beq
\chi(0,{\bf x})=\chi(L,{\bf x})\,\,\,\mbox{and}\,\,\,
\partial_{\mu}\chi(0,{\bf x}) = \partial_{\mu}\chi(L,{\bf
  x})\,.\label{strict-periodic}
\eeq
In addition to these, the theory is also invariant under 
large gauge transformations which are given by
\beq
\chi(0,{\bf x})=\chi(L,{\bf x})+\frac{2\pi}{e}k\,,\,\,\,\,\, k=                                                                                \pm
1,\pm 2,\dots .\label{Z-transf}
\eeq
and represent a global $Z$-symmetry. This symmetry implies the
existence of infinitely many equivalent $Z$-vacua in the theory
(\ref{U(1)-Plus-GN}). Namely, the gauge field
configurations
\beq
A_{\mu}^{Z}(\tau,{\bf x})=(\frac{2\pi}{eL}k,0,0),\,\,\,k=0,\pm
1,\pm 2,\dots, \label{Z-vacua}
\eeq
are lowest energy and equivalent since they are connected by 
gauge transformations of the form $\chi(\tau,{\bfx})=(2\pi/eL)\tau$. 
Still, such $Z$-transformations preserve the anti-periodic boundary 
conditions.  We therefore conclude that
Eq.(\ref{Z-vacua}) represents
equivalent {\it vacuum} configurations of 
(\ref{U(1)-Plus-GN}).

Going back to  (\ref{GN-A-Model}) provided condition
(\ref{Flux-Quantization}) holds, the theory  can be interpreted as
a  Gross-Neveu model with a $U(1)$ gauge coupling at
finite temperature and lying in one of the $Z$-vacua. In this case
there exist no anyon-like excitations in the spectrum. If
condition (\ref{Flux-Quantization}) is relaxed, then
(\ref{GN-A-Model}) corresponds to the gauge theory above
where the gauge field fluctuates around the $Z$-vacua. In the
latter case there exist anyon-like excitations in the spectrum.

We can now construct an order parameter to distinguish the two
cases as follows. Consider translational invariant fluctuations
around the $Z$-vacua in (\ref{U(1)-Plus-GN}) such that
\beq
A_{\mu}=A_{\mu}^{Z}+\theta(\tau)\,,\,\,\,\,\,
\theta(0)=\theta(L).
\label{eq33}
\eeq
Take now the quantity
\beq
\langle {\cal P}_{\tilde{e}}\rangle =  \langle e^{{\rm i}
\tilde{e}\int _0^{L}\rmd\tau
\,\theta(\tau)}\rangle,
\label{Abelian-P}
\eeq
where $\tilde{e}/e\neq$ integer and the average is taken with
respect to the partition function (\ref{U(1)-Plus-GN}). It is not
difficult to see that (\ref{Abelian-P}) is invariant under the
strictly periodic gauge transformations (\ref{strict-periodic}).
It is not invariant, however, under the $Z$-transformations
(\ref{Z-transf}). To see this consider the transformations
\beq
\theta(\tau)\rightarrow \theta(\tau)+\frac{2\pi}{eL}k \,,\,\,\, k=\pm
1,\pm 2,\dots,\label{theta-transf}
\eeq
which are symmetries of the theory
(\ref{U(1)-Plus-GN}) provided there is $Z$-invariance.
Under (\ref{theta-transf})
\beq
\langle{\cal P}_{\tilde{e}}\rangle\rightarrow e^{\rmi 2\pi
k\frac{\tilde{e}}{e}} \langle{\cal P}_{\tilde{e}}\rangle\neq
\langle{\cal P}_{\tilde{e}}\rangle.
\eeq
The physical interpretation of (\ref{Abelian-P}) is that
$F_{\tilde{e}}=-(1/\beta)\ln\langle{\cal P}_{\tilde{e}}\rangle$
gives the free-energy necessary to immerse a configuration of
charge $\tilde{e}$ in the spectrum of the system
\cite{Gross-Pisarski-Jaffe}. If the theory (\ref{U(1)-Plus-GN}) is
in a $Z$-vacuum, then the spectrum can contain only an integer
number of fermions or antifermions since, as explained earlier, it
does not contain anyon-like excitations. In this case one clearly
expects that $F_{\tilde{e}}=\infty$ or $\langle {\cal P}_{\tilde{e}}\rangle=0$. 
If on the other hand the theory is not in a $Z$-vacuum, then there exist 
anyon-like configurations in the spectrum and one in general expects that  
$\langle {\cal P}_{\tilde{e}}\rangle \neq 0$. 
We conclude that, in this sense, the $Z$-symmetry of
the full gauge theory (\ref{U(1)-Plus-GN}) is connected to 
the ``confinement'' or ``screening'' of anyons, 
where $\langle {\cal P}_{\tilde{e}}\rangle$ 
is the relevant order parameter. The anyonic theory preserves $Z$-symmetry in 
the confined phase where $\langle {\cal P}_{\tilde{e}}\rangle =0$ and violates 
the $Z$-symmetry in the deconfined phase where 
$\langle {\cal P}_{\tilde{e}}\rangle \neq 0$.

We are now in a position to give a physical interpretation to the
partition function (\ref{Fint}) with real $\lthetar$ (imaginary
chemical potential), as a result of the following identity
\footnote{Here we use $e=1$.}
\beq
{\cal Z}(T,V,B)=e^{-\beta F(T,V,B)} \equiv \langle {\cal
P}_{\tilde{B}/N}\rangle\,.
\label{Z-as-P}
\eeq
Namely, the free-energy of the Gross-Neveu model with {\it imaginary}
chemical potential represents the free-energy of anyon-like
configurations emerging in the spectrum when the model is coupled
to a $U(1)$ gauge field which fluctuates around the $Z$-vacua.

The above discussion concerning the order parameter
(\ref{Abelian-P}) parallels the discussion in \cite{Semenoff1} of
a similar order parameter for the confinement or screening of
incommensurate charges in parity invariant QED$_{3}$ (see also
\cite{Actor}). One of the new observations here is the
interpretation of the {\it incommensurate charged} configurations
as being anyon-like. Another important point is that the order
parameter $\langle {\cal P}_{\tilde{e}}\rangle$ is in fact
imaginary for any real value of $\tilde{e}$. This is also seen
from (\ref{Constrained-Free-Energy}) or (\ref{Z-as-P}). If
(\ref{Abelian-P}) is to be interpreted as a physical free-energy
density, it is necessary that $\tilde{e}$ is imaginary. The fact
that we obtain an imaginary eigenvalue $B=\tilde{e}$ for the
Hermitian operator $\hat{N}$ means essentially that we are not
using a positive definite density matrix to describe the
fluctuations around the $Z$-vacua of the theory
(\ref{U(1)-Plus-GN}). Nevertheless, even in that case we will
be able to extract useful results for the critical properties of
the theory, assuming that the latter are universal.

\subsection{Chiral Symmetry and $Z$-vacua}

The chiral symmetry restoration for real $\lthetar$ can be
inferred from the gap equation (\ref{gap1}). The critical line
separating the chirally symmetric from the chirally broken phase
in the $\lthetar$-$T$ plane is obtained by setting $\llr=0$ in
(\ref{gap1}) as
\beq
\lthetar = \frac{1}{L}\arccos\left[\frac{e^{Lm}}{2}-1\right],
\eeq
and is depicted for various values of $m$ in Fig.4 when
$\lthetar/T \in [0,\pi]$.

In order to discuss the chiral properties of the theory we shall
henceforth consider, for concreteness, the massless case where the
system is critical already at $T=0$. In this case, when the chemical
potential is real one does not expect any phase transition as the
temperature rises up. To put it differently and in a more general ground, 
for real chemical potential the system is chirally symmetric at some high 
temperature and as it cools down chiral symmetry is, in general,
broken at some lower critical temperature. Having chosen $m=0$ this
critical temperature is $T=0$. Consider now the case when
$\lthetar$ is real, where this corresponds to an imaginary chemical
potential or some fluctuation around the $Z$-vacua as explained
above. Again, at high enough temperatures we expect the
system to be chirally symmetric. However, chiral symmetry is now
broken at a non-zero temperature $T=3\lthetar/2\pi$. The important point 
is that as the system
cools down, chiral symmetry remains broken in a ``temperature
window'' $3\lthetar/2\pi \geq T \geq 3\lthetar/ 4\pi$ and then it
is restored again. Furthermore, as $T\rightarrow 0$, it passes through
infinitely many ``windows'' of the form
\beq
\frac{3\lthetar}{2\pi}\frac{1}{1+3k}\geq T\geq \frac{3\lthetar}{2\pi}
\frac{1}{2+3k} \,, \,\,\, k=0,1,2,\dots,
\label{Theta-Windows}
\eeq
in which chiral symmetry is always broken. This essentially means
that for any real non-zero $\lthetar$ the $T=0$ theory becomes
unstable. The relevant physical picture can be read off from Fig.5
where we plot $\llr/T$ vs $\lthetar/T$ in  $0\leq m/T\leq 2\ln
2$. Note  that for $m/T>2\ln 2$ the theory is always in a chirally
broken phase.

We can draw a physical picture for the chiral symmetry restoration
discussed above. The magnetic flux (\ref{Charge-Flux}) associated with
local charge  fluctuations catalyzes symmetry breaking as
it stabilizes the chiral condensates against thermal fluctuations
inside the ``temperature windows'' (\ref{Theta-Windows})
\footnote{This picture may be compared to the one of an external
  magnetic field in finite temperature 2+1 dimensional QED recently discussed
  in \cite{Metikas}.}. The fact
that in the high-temperature weakly coupled regime even a small
imaginary chemical potential induces chiral symmetry breaking,
shows that the former plays the role of a strong catalyst of
dynamical symmetry breaking, similar to that of a transverse
external magnetic field \cite{Miransky} or a constant negative
curvature \cite{Gorbar}. The chiral symmetry restoration
transition at $T=3\lthetar/2\pi$ is of the second-order with mean
field critical exponents \cite{Huang,Marcello} since the order
parameter $\llr$ is continuous. The order parameter acquires its
maximum value for $T=\lthetar/\pi$ and then starts to drop until
it reaches zero again at $T=3\lthetar/4\pi$. The same picture
holds for all chiral restoration ``temperature windows''
(\ref{Theta-Windows}).

It is also interesting to study the behavior of the free-energy
density as a function of the temperature for non-zero $\lthetar$.
This is depicted in Fig.6 where we plot the free-energy density vs
$\lthetar/T$. For high enough $T$ the system is in the chirally
symmetric phase and to leading-$N$ the free-energy density equals
that of $N$ massless free four-component Dirac fermions e.g.
Eq.(\ref{f-N-Dirac}). At $T=3\lthetar/2\pi$ the free-energy jumps
discontinuously to a local minimum and then raises until it
reaches a local  (positive) maximum at $T=3\lthetar/4\pi$ where it
drops again discontinuously to the value (\ref{f-N-Dirac}).
Similar fluctuations occur for all the infinitely many temperature
``windows'' (\ref{Theta-Windows}) as $T\rightarrow 0$.

It is rather intriguing that we are able to give analytic expressions
for the free-energy density of the system at the end-points of the
``temperature windows'' of chiral symmetry breaking, $3\lthetar/2\pi$,
$3\lthetar/4\pi$ (in fact this is possible for the end-point of all
the ``windows'' (\ref{Theta-Windows})). After some algebra
we obtain respectively
\bea
\frac{f_{L}}{N}  &=&  -\frac{1}{\pi
L^3}\left[\frac{4\pi}{3}Cl_2\left(\frac{\pi}{3}\right)
-\frac{2}{3}\zeta(3)\right]\,,
\label{f-one-edge}\\
\frac{f_{L}}{N}  &=&  \frac{1}{\pi
L^3}\left[\frac{8\pi}{3}Cl_2\left(\frac{\pi}{3}\right)
+\frac{2}{3}\zeta(3)\right]\,,
\label{f-other-edge}
\eea
where, $Cl_2(\phi)={\rm Im}[Li_{2}(e^{{\rm i}\phi})]$ is the
Clausen function \cite{Lewin}. Note that $Cl_2(\pi/3)$ is the
absolute maximum of this function which is a well-documented
irrational number \cite{Kolbig}.

The results for the free-energy density have been on purpose
written as above, in order to be compared with the expected
scaling form of the free-energy density of a conformal field
theory (CFT). The point is that the chiral phase transition in the
2+1 dimensional model above is of  second-order and one would
expect it to correspond to the universality class of a
3-dimensional CFT. When a CFT is put in a finite-size geometry
(i.e. in a slab with one finite dimension of length $L$), its
free-energy density scales as \cite{Cardy}
\beq f_{L}=-\frac{\tilde{c}\,\zeta(3)}{2\pi L^3} \, \cdot
\label{f-scaling} \eeq
The quantity $\tilde{c}$ coincides with the central charge in
$d=2$ and has been recently proposed \cite{Appelquist} to be  a
possible generalization of a $Cl$-function in $d>2$. Clearly, from
(\ref{f-one-edge}), (\ref{f-other-edge}) and (\ref{f-scaling}) we
see that the chiral transition above appears to be connected to
new 3-dimensional CFTs. The fact that $\tilde{c}$ in
(\ref{f-one-edge}) is less than the corresponding free-field
theory value (\ref{f-N-Dirac}) and it is negative in
(\ref{f-other-edge}), seems to imply that the above critical
theories may not be unitary. Nevertheless, such theories may
conceivably correspond to three-dimensional versions of the
non-unitary two-dimensional Lee-Yang model \cite{Cardy2}.

The middle point of the chiral transition ``temperature window''
(\ref{Theta-Windows}) is also interesting. At this point $\llr$
takes its maximal value. From (\ref{gap1}) we obtain for $\lthetar
=\pi/L$,
\beq
\llr =\frac{2}{L}\ln \left(\frac{1+\sqrt{5}}{2}\right).
\eeq
This value equals the value one obtains for the mass of the
elementary bosonic modes in the 2+1 dimensional $O(N)$ vector
model at finite temperature, when the theory is critical at $T=0$
\cite{Sachdev}. Plugging this into (\ref{Constrained-Free-Energy})
we can calculate the free-energy density which, by virtue of some
non-trivial polylogarithmic identities \cite{Sachdev,Tassos2} is
found to be
\beq
\frac{f_{L}}{N} = \frac{16}{5}\frac{\zeta(3)}{2\pi L^3}.
\label{f-center}
\eeq
This is exactly {\it minus} the free-energy density of the 2+1
dimensional $O(4N)$ vector model at its non-trivial critical
point. In principle, a positive free-energy density which would
correspond to negative pressure and negative entropy seems to be a
rather unphysical result \cite{Smilga}. However, one might try to
construct a physical system where the result (\ref{f-center})
could make sense. This is the $O(4N)$ ${\cal N}=1$ supersymmetric
sigma model \cite{Gracey,Koures} in the presence of an external
$A_{0}$ potential at finite temperature.  To leading-$N$, the
free-energy density of this theory is simply given by the sum of
the free-energies of fermions and bosons. In general,
supersymmetry is expected to be broken at any finite temperature
\cite{Boyanovsky}. Nevertheless, it may happen that for some
temperature the fermion contribution, given by (\ref{f-center}),
and the boson contribution cancel each other and the system
becomes supersymmetric again. Note that the matching of the
bosonic and fermionic degrees of freedom is correct  - the ${\cal
N}=1$ supermultiplet in three-dimensions requires $2N$
two-component Majorana fermions. A similar bosonization of
fermions has been recently discussed in \cite{Pisarski}.

\section{Summary and Discussion}

In this work we studied  the 2+1 dimensional Gross-Neveu model in
the presence of an imaginary chemical potential and argued that it
provides an interesting ground for probing the properties of
chiral symmetry at non-zero temperature. In Sec.3 we proposed  a
generalization of the well-known formula for the  canonical
partition function and presented analytic expressions for the
free-energy density, the fermion number density and the chiral
order parameter $\llr$. We demonstrated that this general
formalism includes the standard  grand-canonical formalism
when the chemical potential is real. In Sec.4 we focused on the
case of imaginary chemical potential. We considered a $U(1)$ gauge
field coupled to the Gross-Neveu model at finite temperature.
Such a theory possesses infinitely many equivalent $Z$-vacua. 
We showed that when the gauge field fluctuates around these Z-vacua, 
as given by Eq.(\ref{eq33}), anyon-like excitations manifest 
themselves in the spectrum of the 2+1 dimensional Gross-Neveu model. 
This is established by showing that the free-energy necessary to 
immerse such an excitation is non-zero. Furthermore, we provided evidence 
that the expectation value of the abelian Polyakov loop, which is the order
parameter for establishing whether the above $U(1)$ gauge theory
resides in one of the $Z$-vacua, coincides with the {\it
canonical} partition function for imaginary values of the fermion
number density. This way we gave a physical interpretation to the
 canonical partition function with imaginary chemical
potential as being the free-energy needed to immerse an excitation
of imaginary charge in the spectrum of the 2+1 dimensional
Gross-Neveu model. The latter excitation is related to the
anyon-like excitations discussed above. Finally, we gave some
results connected to the properties of chiral symmetry in the
presence of an imaginary chemical potential at finite temperature.

Studies of the critical thermodynamic properties of a system
undergoing a phase transition are intimately connected to the
Lee-Yang zeroes \cite{Lee-Yang,Barbour}. These are the zeroes of
the partition function for imaginary values of the external
magnetic field, the latter being the ``conjugate'' variable of the
relevant order parameter. It is then conceivable that one could
try to investigate the critical properties of a system at finite
density by studying its partition function (or its free-energy
density), at complex values of the number density, the latter
being the ``conjugate'' variable of the chemical potential. In
this sense, the results presented in this work are closely related
to a Lee-Yang zeroes analysis.

These results can be extended in several directions. For
example,it is possible in principle to study numerically the
partition function (\ref{Fint}) for complex chemical potential and
establish the chiral symmetry restoration ``windows'' advocated above. 
It is also possible to numerically compute the free-energy densities at
the critical points and compare them with our analytic results
(\ref{f-one-edge}) and (\ref{f-other-edge}). Furthermore, it would
be interesting to compare the free energy density
(\ref{Constrained-Free-Energy}) to the one recently proposed by
Laughlin \cite{Laughlin} in order to reproduce the anomalous
behaviour of the thermal conductivity in some high-temperature
superconductors of the BSCCO family \cite{Krishana} \footnote{See
also \cite{Semenoff-Mavromatos-Liu}.}.
Since these compounds exhibit a new kind of phase transition
induced by strong catalysts of dynamical symmetry breaking, it is
plausible that our generalized canonical partition function
formalism can be used to explain such unusual behavior. Finally, based on 
the results of the present work, it may be interesting to explore the
possibility of finite temperature supersymmetry restoration in
${\cal N}=1$ supersymmetric theories \cite{mavromatos}.

\noindent
{\bf Acknowledgements}

The work of H.R.C. and M.B.S.N was supported in part by the
Brazilian agencies for the development of science FAPERJ and
CAPES. A.C.P was supported in part by an I.K.Y. Postdoctoral
Fellowship and an Alexander von Humboldt Fellowship. N.D.V. was
supported by the E.U. under TMR contract No. ERBFMRX--CT96--0090.














\begin{figure}[ht] 
\epsfxsize=14cm
\centerline{\epsffile{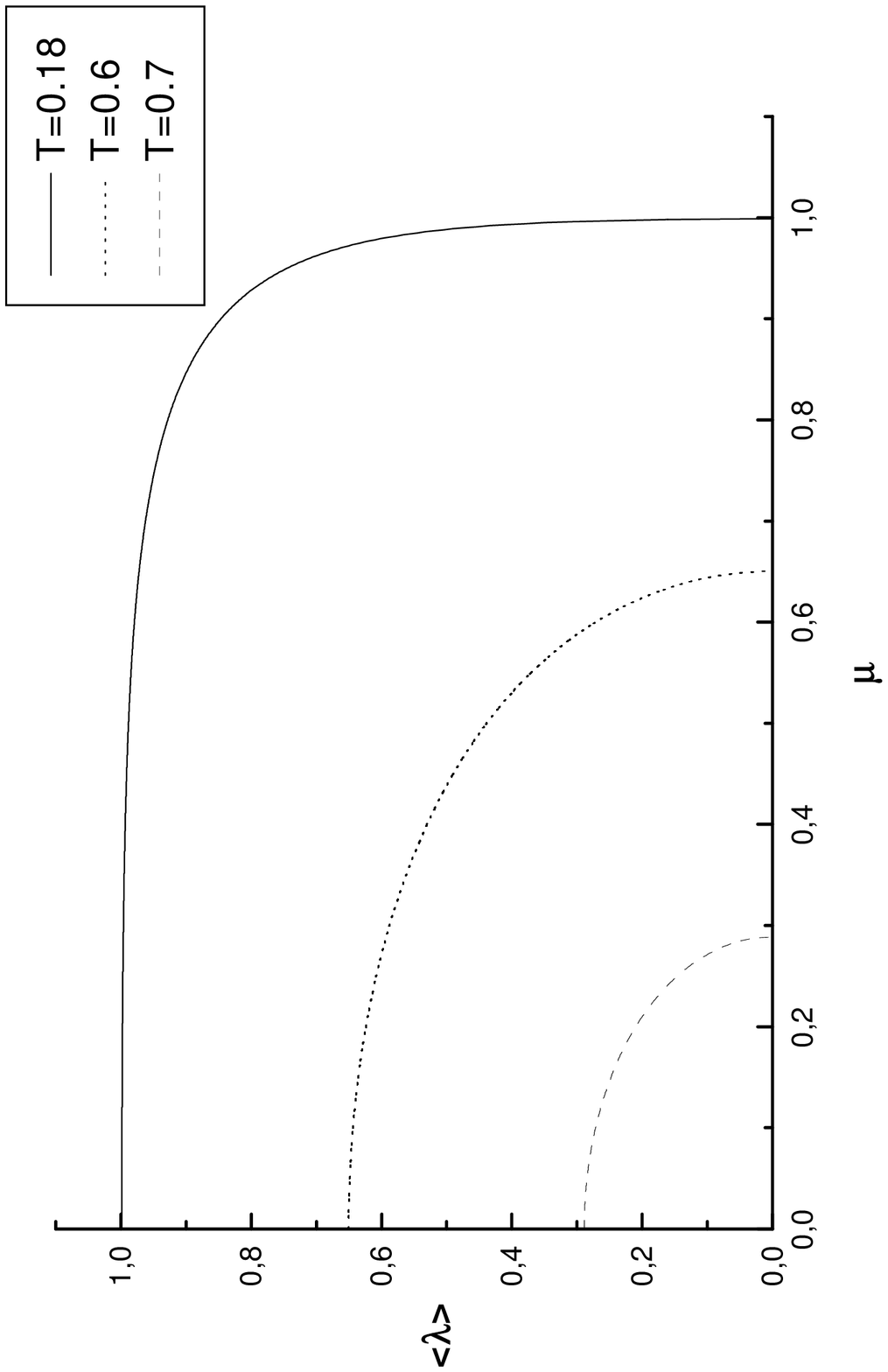}}
\caption{Critical lines in the $\llr - \mu$ plane for various
  $T<T_{c}$. For simplicity we chose $m=1$ such that $T_{c}\approx 0.72$.}
\end{figure}


\begin{figure}[ht]
\centerline{\epsfxsize=14cm \epsffile{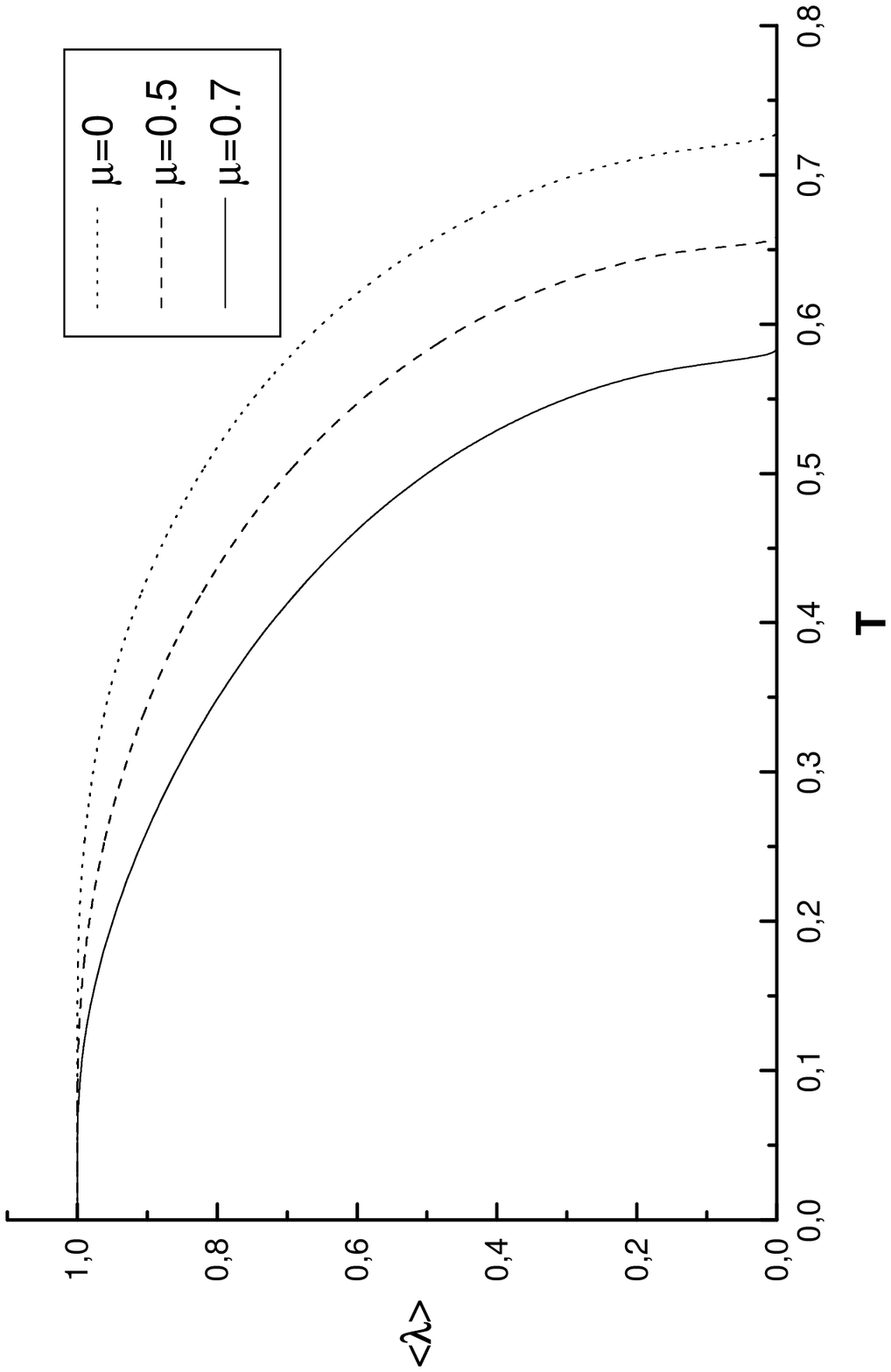}}
\caption{$\llr$ vs $T$ for various $\mu$, when $m=1$.}
\end{figure}


\begin{figure}[ht]
\centerline{\epsfxsize=14cm \epsffile{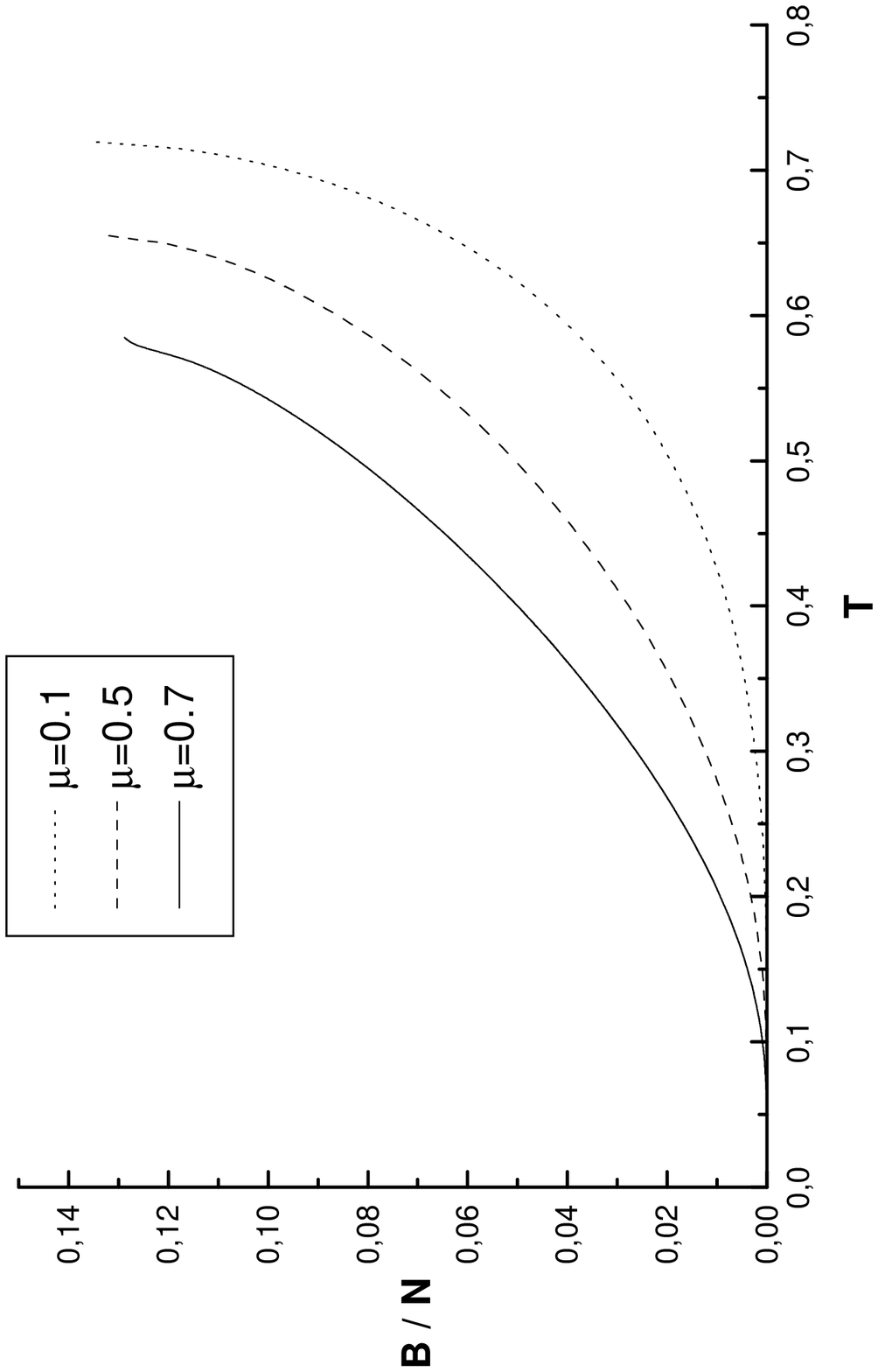}}
\caption{$\tilde{B}/N$ vs $T$ for various $\mu$ and $m=1$. The baryon
  density blows up at $T\rightarrow T_{c}(\mu)$.}
\end{figure}


\begin{figure}[ht]
\centerline{\epsfxsize=14cm \epsffile{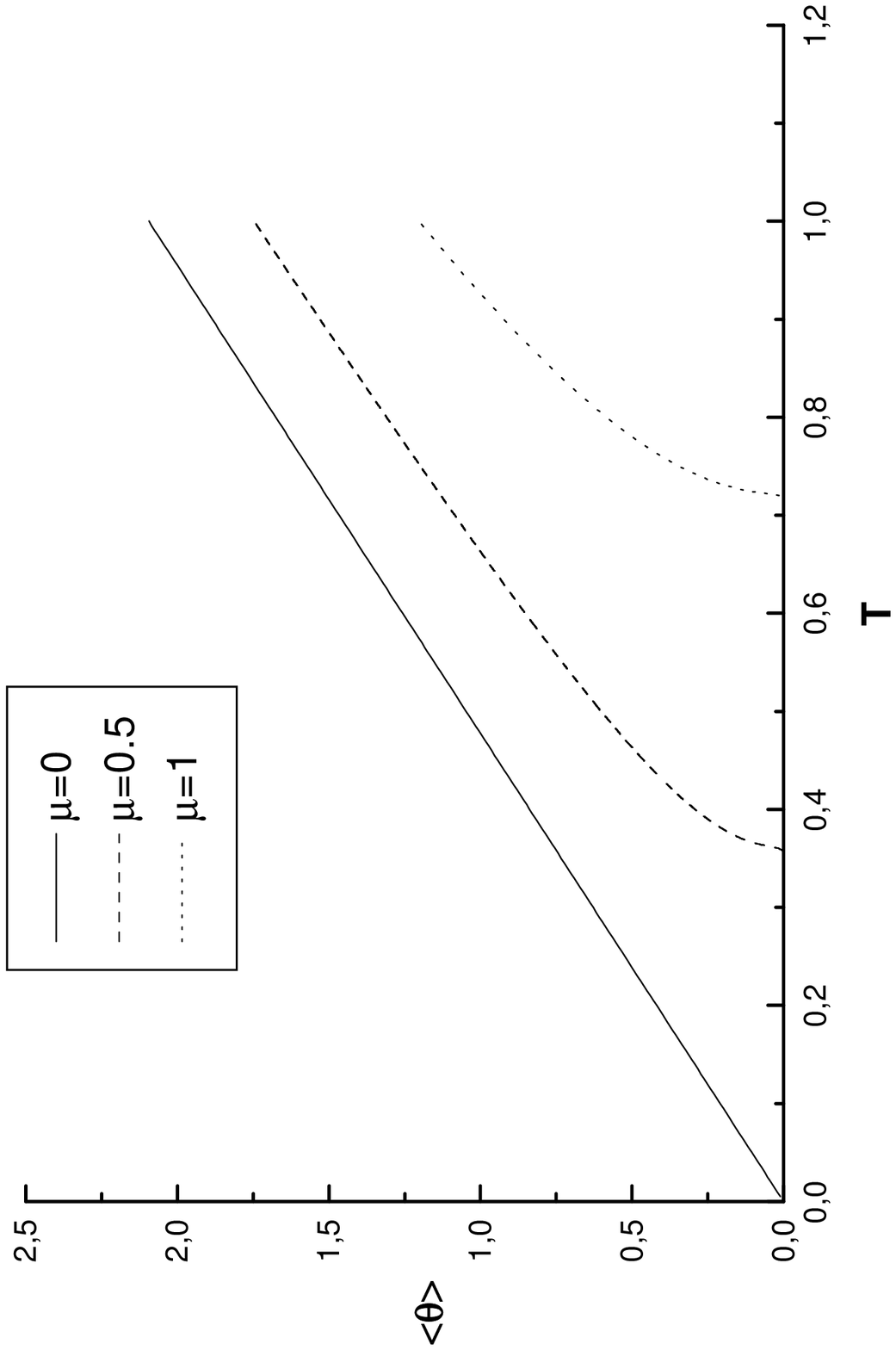}}
\caption{Critical lines in the $\lthetar - T$ plane for various $m$ when
  $\lthetar/T\in [0,\pi]$. The line $\lthetar=(2\pi/3)T$ is an asymptote.}
\end{figure}


\begin{figure}[ht]
\centerline{\epsfxsize=14cm \epsffile{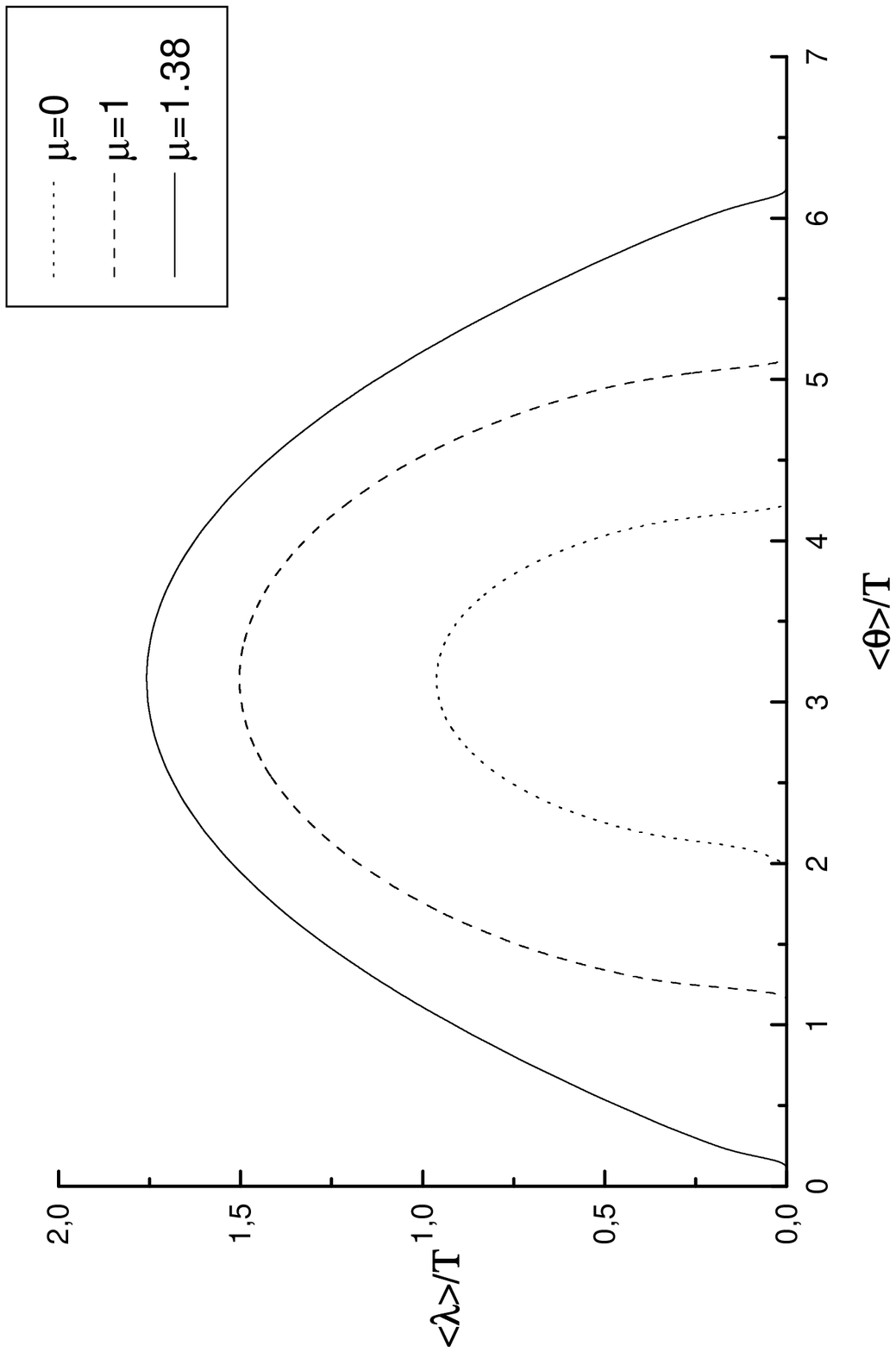}}
\caption{$\llr /T$ vs $\lthetar /T$ for $0<m/T <2\ln 2$. For
  fixed, non-zero $\lthetar$, $T$ decreases along the horizontal axis
  as one  moves to the right.}
\end{figure}


\begin{figure}[ht]
\centerline{\epsfxsize=14cm \epsffile{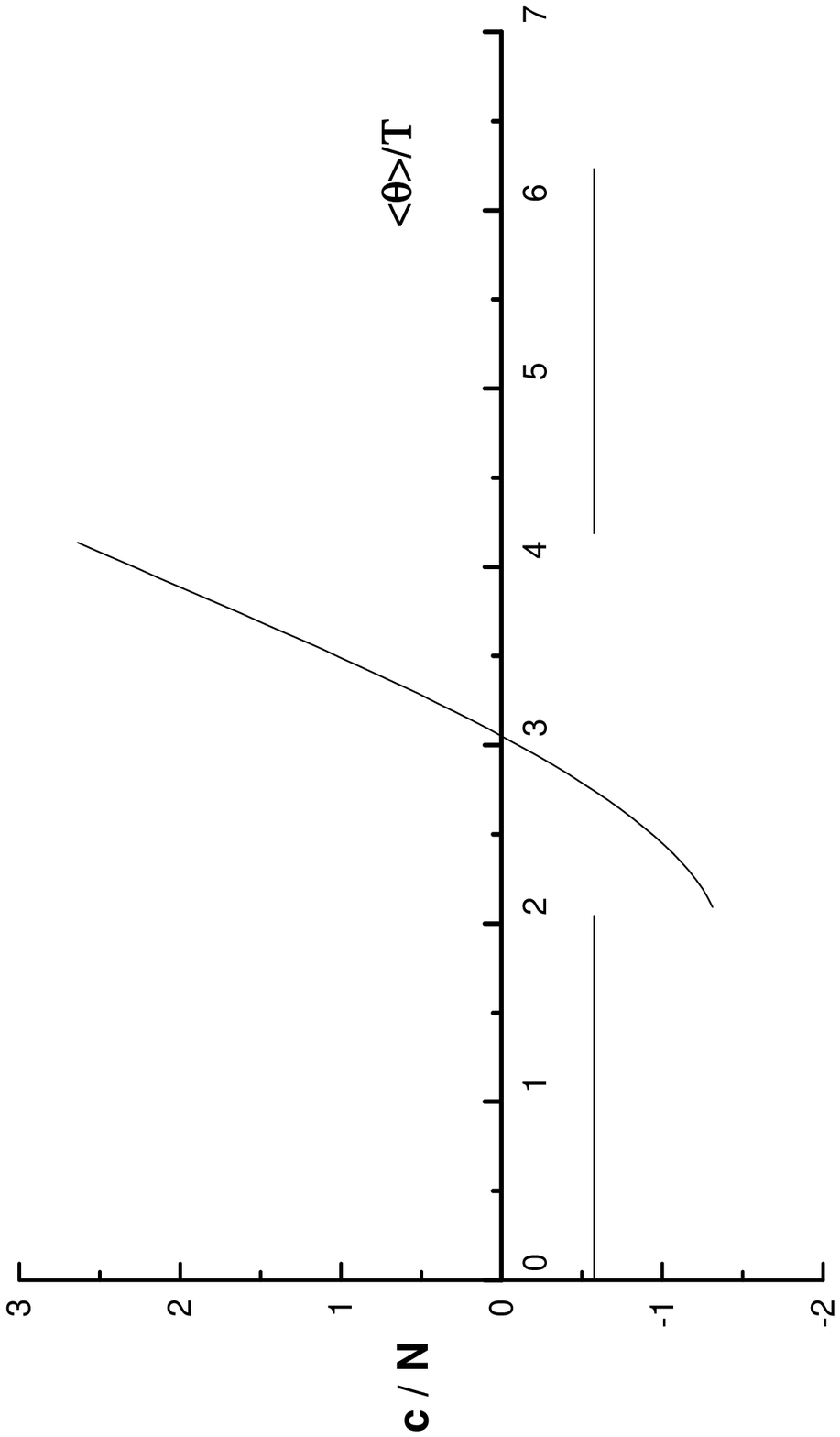}}
\caption{The free-energy density vs $\lthetar /T$. As in Fig.5, $T$
  decreases along the horizontal axis as one moves to the right.}
\end{figure}

\end{document}